\documentclass{aa}

\usepackage{graphicx}
\usepackage{txfonts}
\usepackage[colorlinks=true, linkcolor=blue, citecolor=blue,
            filecolor=blue, urlcolor=blue]{hyperref}
\usepackage{subcaption}
\usepackage{multirow}
\usepackage{arydshln}
\usepackage{booktabs}
\usepackage{amsmath}

\graphicspath{{figures/}}

\DeclareMathOperator*{\argmax}{argmax}

\begin{document}

\title{The joint wavelet $\ell_1$-norm matches neural compression
       for tomographic weak-lensing inference}

\author{%
    Andreas Tersenov \inst{1,2,3,4,5}\fnmsep\thanks{\email{atersenov@physics.uoc.gr}}
    \and
    Jean-Luc Starck \inst{2,5}
    \and
    Martin Kilbinger \inst{5}
    }

\institute{
            Department of Physics, University of Crete, Greece
            \and
             Institutes of Computer Science and Astrophysics, Foundation for Research and Technology Hellas (FORTH), Greece
            \and 
             Department of Astronomy, University of Virginia, 530 McCormick Road, Charlottesville, VA 22904, USA
            \and
             Virginia Institute for Theoretical Astronomy, University of Virginia, Charlottesville, VA 22904, USA
            \and
             Universit\'e Paris-Saclay, Universit\'e Paris Cit\'e, CEA, CNRS, AIM, 91191, Gif-sur-Yvette, France
   }


%
\abstract
{Much of the cosmological information in Stage~IV weak-lensing surveys lies on non-linear scales, where
two-point statistics capture only part of it. Learned neural compressors are presented as near-optimal
summaries of convergence maps, yet rarely tested against a strong hand-crafted statistic under
controlled conditions.
}
{We investigate whether a convolutional compressor extracts more information from tomographic convergence
maps than the starlet $\ell_1$-norm, how the $\ell_1$-norm can be extended to capture the
correlations between redshift bins, and whether the weaker constraints that higher-order statistics
give after the Bernardeau--Nishimichi--Taruya (BNT) nulling transform reflect a genuine loss of
information.
}
{We perform simulation-based inference on the \texttt{cosmoGRID} simulations. We introduce two extensions of the $\ell_1$-norm to pairs of bins. The first adds the
pixel-wise products of pairs of maps as extra channels. The second, the joint
$\ell_1$-norm, is an $\ell_1$-weighted joint histogram of the wavelet coefficients of each bin pair.
}
{The analytical statistic is near-sufficient: with the product cross-maps, the
$\ell_1$-norm comes within $10\%$ of the compressor's figure of merit, and the joint
$\ell_1$-norm reaches it within the uncertainties. The BNT transform is invertible, so no information is lost. Individually,
however, the nulled maps carry little signal and their noise is correlated, so the information can
be recovered only from the maps jointly. A per-bin statistic therefore loses most of its
constraining power, the joint $\ell_1$-norm keeps most of it, and the CNN is near-lossless.
}
{A fixed wavelet statistic that uses the tomographic maps jointly is as constraining as an
optimised neural compressor. The network's remaining advantage is that it combines the maps by construction, so it
needs no cross-maps and is least affected by the nulling. The
$\ell_1$-norm, in turn, requires no training, and its data vector can be inspected.
}


\titlerunning{The joint wavelet $\ell_1$-norm matches neural compression}
\authorrunning{Tersenov et al.}

\maketitle
\nolinenumbers

\section{Introduction}
\label{sec:intro}

The Stage~IV weak-lensing surveys, \textit{Euclid} \citep{euclid2024}, the Vera~C.~Rubin Observatory
\citep{ivezic_lsst_2019} and the Roman Space Telescope \citep{spergel_roman_2015}, will measure the
gravitational distortion of galaxy shapes over thousands of square degrees and map the total matter
distribution \citep{bartelmann_weak_2001, kilbinger_cosmology_2015, mandelbaum_weak_2018} with a
statistical power well beyond that of the current KiDS, DES, HSC and UNIONS surveys
\citep{kuijken_KiDS_2019, flaugher_dark_2005, aihara_HSC_2017, hervas-peters_unions_2026}. Much of
the information lies on small, non-linear scales, where the convergence field is strongly
non-Gaussian and is therefore only partly characterised by its two-point statistics
\citep{weinberg_observational_2013}. Higher-order statistics have been developed to extract the
remaining information: the bispectrum \citep{Takada_three-point_2003, takada_cosmological_2004},
Minkowski functionals \citep{kratochvil_probing_2012, parroni_higher-order_2021}, peak counts,
which probe the abundance of massive haloes \citep{Marian_cosmology_2009, dietrich_cosmology_2010,
lin2015new, martinet_probing_2021, aycoberry_unions_2023}, the scattering transform
\citep{cheng_new_2020, valogiannis_towards_2022, cheng_weak_2021} and the starlet 
$\ell_1$-norm \citep{ajani_starlet_2021, ajani_starlet_2023}, which summarises the full one-point
distribution of the wavelet coefficients across scales and has been shown to be significantly more
constraining than the power spectrum in Euclid-like forecasts. The $\ell_1$-norm also admits a
theoretical prediction, derived within large-deviation theory from the one-point distribution of
the convergence, which reproduces simulations accurately in the weakly non-linear regime
\citep{sreekanth_theoretical_2024}.

Each of these statistics is a fixed, hand-crafted function of the maps, designed to be sensitive to
a particular feature of the field. The alternative is to learn the compression, training a neural
network to map the convergence field to a low-dimensional summary that is maximally informative
about the cosmological parameters \citep{fluri_full_2022, jeffrey_dark_2024, omori_towards_2026}.
Such learned compressors are often presented as near-optimal, extracting essentially all the
cosmological information the maps contain. Their constraining power, however, is usually assessed
by comparison with two-point statistics, which many non-Gaussian summaries already surpass, and
rarely by comparison with a strong hand-crafted higher-order statistic under matched conditions.
We investigate here whether a hand-crafted statistic can extract as much cosmological information
from the maps as an optimised learned compression. Both summaries enter the same inference
framework, simulation-based inference (SBI), which learns the parameter posterior from forward
simulations without an explicit likelihood \citep{cranmer_frontier_2020} and represents it with a
conditional normalising flow \citep{papamakarios_masked_2017}.

Tomography adds a further consideration. Splitting the sources into redshift bins places
information both in the convergence within each bin and in the correlations between bins, which
constrain the growth of structure over cosmic time and the distance--redshift relation. A
statistic evaluated on each bin separately, for example the $\ell_1$-norm or peak counts in their
standard form, captures only the former. Explicit tomographic cross-maps have been proposed to
supply the information between bins as additional inputs \citep{zurcher2022dark,
zurcher_towards_2023}. The Bernardeau--Nishimichi--Taruya transform
\citep[BNT;][]{bernardeau_cosmic_2014}, a linear recombination of the bins into maps whose lensing
kernels are localised in redshift (the nulled maps), could serve the same purpose in a different way. Since the nulled
kernels are compact and do not overlap, the signals of the nulled maps are largely uncorrelated,
and the information that the original maps hold in their correlations is then largely contained
in the individual nulled maps, where a statistic evaluated on each map separately can capture it.
The transform also confines any systematic that is localised in redshift, such as baryonic
feedback at low redshift and small scales \citep{van_daalen_effects_2011,
semboloni_quantifying_2011, chisari_modelling_2019}, to a few nulled maps and scales, where it can
be removed by a scale cut \citep{tersenov_baryonic_2026}. Applied before map-based higher-order
statistics such as the $\ell_1$-norm or peak counts, however, the transform has been found to
inflate the cosmological contours \citep{tersenov_baryonic_2026, vinciguerra_euclid_2026}. A
learned compressor that takes the tomographic maps as joint input might be expected to capture a
larger part of the information between bins than a statistic evaluated on each bin separately.

In this work we compare the starlet $\ell_1$-norm with a convolutional compressor trained by
variational mutual-information maximisation, for tomographic weak-lensing inference on the
\texttt{cosmoGRID\,V1} simulations \citep{kacprzak_cosmogridv1_2023, fluri_full_2022}. The
inference is optimised for each summary independently, both summaries are compressed to the same
dimension and modelled by the same density estimator, and the calibration of every posterior is
verified. We aim specifically to:
\begin{enumerate}
  \item compare the constraining power of the hand-crafted $\ell_1$-norm and the learned
        compressor in terms of the marginal uncertainties and a three-parameter figure of merit;
  \item establish how best to capture the inter-bin correlations in a map-based higher-order
        analysis and quantify the gain in constraining power;
  \item determine whether the BNT contour inflation of the $\ell_1$-norm reflects
        information that is genuinely lost or only the basis in which the statistic is evaluated,
        and whether a compressor that uses the maps jointly is affected.
\end{enumerate}
The paper is organised as follows. Section~\ref{sec:simulations} describes the
\texttt{cosmoGRID\,V1} simulations and the tomographic flat-sky patches. Section~\ref{sec:stats}
defines the two compression strategies and the tomographic cross-maps. Section~\ref{sec:inference}
sets out the inference and the calibration tests. Section~\ref{sec:results} compares the
constraining power of the two summaries. Section~\ref{sec:bnt} turns to the BNT transform and
shows the contour inflation of the $\ell_1$-norm to be an artefact of the analysis basis.
Section~\ref{sec:discussion} discusses what the cross-maps contribute, why the $\ell_1$-norm can be
as constraining as a learned compressor, the practical differences between the two summaries, the
use of the nulled basis, and the limitations of the comparison. We conclude in
Section~\ref{sec:conclusions}.
\section{Simulations and data}
\label{sec:simulations}

We base our analysis on the \texttt{cosmoGRID\,V1} suite \citep{kacprzak_cosmogridv1_2023, fluri_full_2022},
a set of full-sky weak-lensing convergence maps produced with the \texttt{PkdGrav3} $N$-body code
and provided at a HEALPix \citep{gorski_healpix_2005} resolution of $N_{\rm side}=512$, sufficient
to probe multipoles $\ell\le1024$. The maps span a flat $w$CDM model in six parameters,
$\boldsymbol{\theta}=(\Omega_{\rm m},\sigma_8,w_0,H_0,n_{\rm s},\Omega_{\rm b})$, sampled with a Sobol
sequence. The fiducial cosmology is $\Omega_{\rm m}=0.26$, $\sigma_8=0.84$, $w_0=-1$. We report
constraints in the subspace $(\Omega_{\rm m},\sigma_8,w_0)$, since the remaining parameters are only
weakly constrained by weak lensing. The maps are provided in four tomographic redshift bins, with
mean redshifts $z\approx0.3,0.5,0.7,0.9$. We add Gaussian shape noise at the map level for a
\textit{Euclid}-like configuration, with an effective source density
$n_{\rm gal}=30\,\mathrm{arcmin}^{-2}$ split across the four bins and an intrinsic ellipticity
dispersion $\sigma_\epsilon=0.26$ per component. Throughout, we work directly on the convergence
maps, with no mass-mapping reconstruction from shear.

We analyse the maps as flat-sky patches, the natural input for a convolutional compressor. We
extract non-overlapping $10\degr\times10\degr$ gnomonic patches of $80\times80$ pixels (a pixel
scale of $7\farcm5$), restricted to latitudes $|b|<75\degr$ to avoid the projection distortion of
the polar caps. This gives $180$ patches per realisation. We chose $10\degr$ patches for the
accuracy of the flat-sky approximation. The fractional distortion of the gnomonic projection at the
patch corners is $1.5\%$. For the per-patch population analysis of
Sect.~\ref{sec:results} we draw $50$ independent noise realisations of each patch, giving
$180\times50=9000$ mock observations per configuration. The learned stages of the analysis (the
compressors and density estimators of Sects.~\ref{sec:stats} and~\ref{sec:inference}) are trained
on $3.2\times10^5$ patch examples drawn from $899$ of these cosmologies, the same set for both
summaries.

Finally, we subtract the mean of each map before compression, since the mean convergence over a
finite field is not an observable, owing to the mass-sheet degeneracy, and retaining it would
favour the learned compressor (Appendix~\ref{app:pitfalls}).
\section{Summary statistics and compression}
\label{sec:stats}

\begin{figure*}[t]
  \centering
  \includegraphics[width=0.9\textwidth]{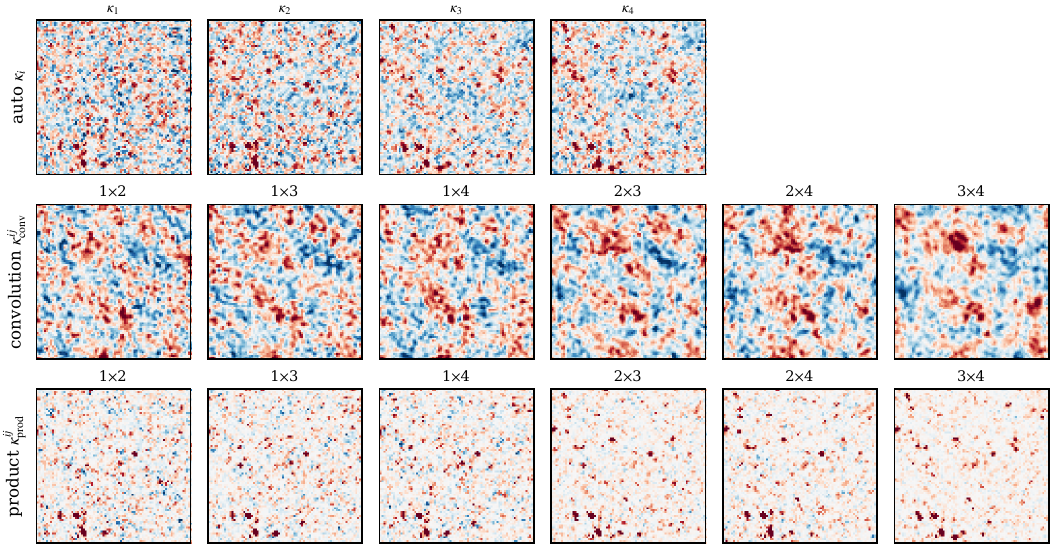}
  \caption{The input channels for a representative $10\degr\times10\degr$ patch: the four tomographic
  auto-maps $\kappa_i$ (top), and the two cross-maps of Sect.~\ref{sec:crossmaps} for each bin pair
  $i<j$, the convolution $\kappa^{ij}_{\rm conv}$ (middle) and the pointwise product
  $\kappa^{ij}_{\rm prod}$ (bottom). The cross-maps are computed from the auto-maps of the patch. The
  $\ell_1$-norm is evaluated on each channel separately, whereas the CNN takes the four auto-maps as
  joint input.}
  \label{fig:data}
\end{figure*}

We compare two strategies for compressing the tomographic convergence maps into a low-dimensional
summary for the inference of Sect.~\ref{sec:inference}: a hand-crafted multiscale statistic, the
starlet $\ell_1$-norm, and a learned compressor, a convolutional neural network (CNN) trained with
the variational mutual-information maximisation (VMIM) objective. Throughout, we call the first
summary analytical in the sense that it is a fixed, closed-form function of the maps with no
trained component. Both start from the same four tomographic convergence maps $\kappa_i$
($i=1,\dots,4$). We call the maps that enter a summary its \emph{channels}. The CNN takes the four
auto-maps together as its input and combines them from its first convolutional layer
(Sect.~\ref{sec:cnn}). The $\ell_1$-norm, in contrast, is evaluated on each channel individually, so
inter-bin information must be supplied to it by explicit constructions.
Two such constructions appear below: the tomographic \emph{cross-maps}, extra input channels of two types built
for each bin pair (Sect.~\ref{sec:crossmaps}), and the joint $\ell_1$-norm, a pairwise
generalisation of the statistic itself (Sect.~\ref{sec:jointl1}). Figure~\ref{fig:data} shows the
channels for a representative $10\degr$ patch: the four auto-maps (top row), and the cross-maps
(middle and bottom rows).

\subsection{The starlet $\ell_1$-norm}
\label{sec:l1}

\begin{figure*}[t]
  \centering
  \includegraphics[width=\textwidth]{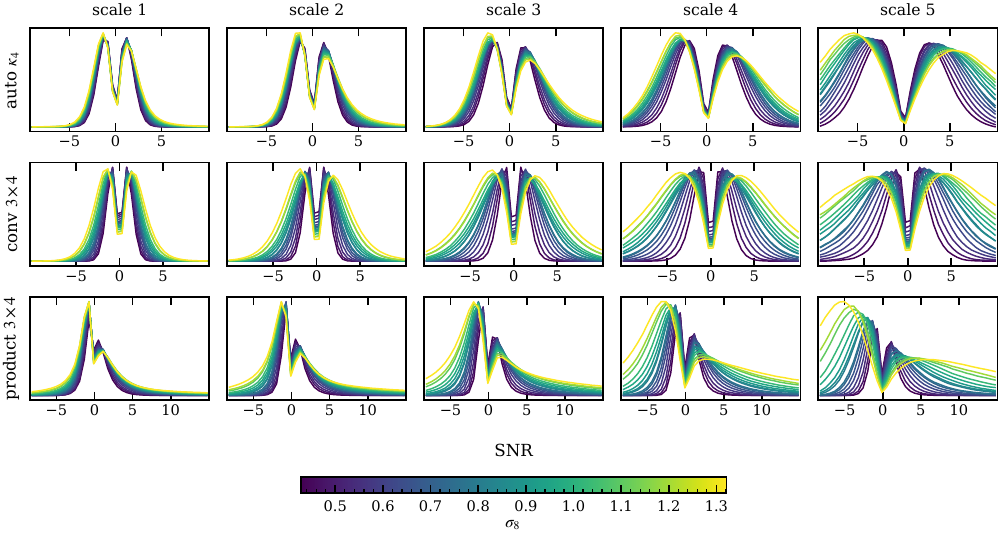}
  \caption{The starlet $\ell_1$-norm summary and its sensitivity to cosmology. For three
  representative channels (an auto-map and two cross-maps), the $\ell_1$ histogram is shown per
  scale (columns) against SNR, with the curves coloured by $\sigma_8$ across sixteen
  equal-count cosmology levels. The histograms shift systematically with $\sigma_8$ at every scale
  and in every channel.}
  \label{fig:l1}
\end{figure*}

The starlet $\ell_1$-norm is a multiscale characterisation of the full probability density of the
convergence field, obtained by summing the absolute values of the starlet wavelet coefficients
within signal-to-noise (SNR) bins. We summarise the construction here and refer to
\citet{ajani_starlet_2021} and \citet{tersenov_baryonic_2026} for the full formalism. The
convergence map is decomposed with the isotropic undecimated (``starlet'') wavelet transform
\citep{starck:sta06} into four wavelet scales and a coarse scale. The wavelet scales are dyadic
bands with characteristic angular sizes of roughly $10\arcmin$, $20\arcmin$, $40\arcmin$ and
$80\arcmin$, and the coarse scale holds the remaining large-scale signal. The statistic is computed
on all five scales (Fig.~\ref{fig:l1}). For each scale we bin the wavelet coefficients by their SNR, the
coefficient divided by the noise level of that scale, into $40$ linearly spaced SNR bins and sum
the absolute wavelet coefficients in each bin. Concatenating across SNR bins and scales gives the
data vector.

We compute the $\ell_1$-norm on the auto-maps and, in the configurations of
Sect.~\ref{sec:crossmaps}, on the cross-maps as well. Because the cross-maps are smaller in
amplitude than the auto-maps by orders of magnitude, we set the SNR binning per channel, with a range of $[-13,13]$ for the auto
channels and the cross channels rescaled by their own fixed noise estimate before the wavelet transform
(Appendix~\ref{app:pitfalls}). We then apply a $\log(1+x)$ transform, standardisation, and clipping
at five standard deviations to the data vector.

\subsection{The joint $\ell_1$-norm}
\label{sec:jointl1}

\begin{figure}[t]
  \centering
  \includegraphics[width=0.8\columnwidth]{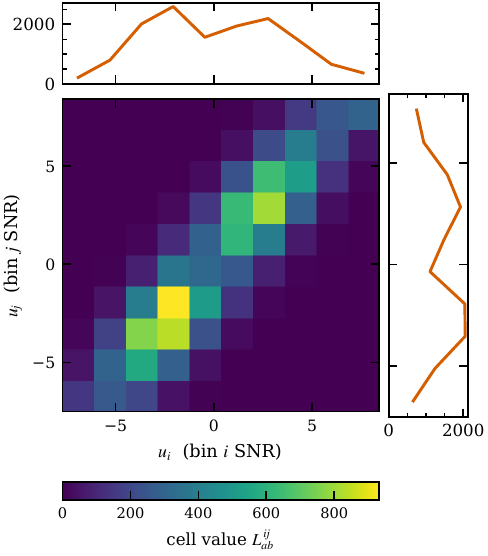}
  \caption{The joint $\ell_1$-norm for one pair of maps and one wavelet scale. Each cell holds the
  summed $\ell_1$ weight of the pixels whose coefficient pair $(u_i,u_j)$ falls in it,
  Eq.~\eqref{eq:jointl1}. The side panels show the sums of the cells along each axis, which
  correspond to the per-map $\ell_1$-norms of the two maps.}
  \label{fig:l1joint}
\end{figure}

At each wavelet scale, let $u_i(\boldsymbol{p}) = (W\kappa_i)(\boldsymbol{p})/\sigma_i$ be the
wavelet coefficient of map $i$ at spatial pixel $\boldsymbol{p}$ in SNR units, where $\sigma_i$ is
the noise level of the wavelet coefficients of map $i$ at that scale. The per-channel $\ell_1$-norm
of Sect.~\ref{sec:l1} bins these coefficients by their SNR value, and each SNR bin $B$ holds the
sum of the absolute wavelet coefficients that fall in it, $\sum_{u_i\in B}|u_i|$. The joint
$\ell_1$-norm is the two-dimensional version of this histogram for a pair of maps $(i,j)$. Every
pixel gives one pair $(u_i,u_j)$, formed from the coefficients of the two maps at that same pixel.
We divide the $(u_i,u_j)$ plane into a fixed grid of $K\times K$ cells, with $K$ SNR bins
$B_1,\dots,B_K$ along each axis, so that cell $(a,b)$ is the square $B_a\times B_b$. Each cell
holds the sum, over the pairs that fall in it, of the average of the two absolute wavelet
coefficients,
\begin{equation}
\label{eq:jointl1}
L^{ij}_{ab} = \sum_{(u_i,\,u_j)\,\in\,B_a\times B_b} \tfrac{1}{2}\big(|u_i| + |u_j|\big).
\end{equation}
Cells on the diagonal, $a=b$, thus collect the pixels at which the two maps have the same SNR, and
cells away from it those at which they differ. Figure~\ref{fig:l1joint} shows the resulting array
of cell values for one pair of maps and one wavelet scale. The data vector concatenates the $K^2$
cell values over the six pairs of maps and the wavelet scales. We use $K=10$ and, as for the other
summaries, compress the data vector to ten dimensions before inference. The joint $\ell_1$-norm is
computed from the four auto-maps alone, without any cross-map, and captures how the wavelet
coefficients of two maps vary together pixel by pixel.

In principle the same histogram could be built over three or all four maps at once, but in
practice the number of cells grows too fast for the pixels of a single patch to populate it. We
therefore stop at pairs.

\subsection{The CNN-VMIM compressor}
\label{sec:cnn}

In contrast to the fixed $\ell_1$-norm, a learned compressor optimises its summary directly for
the inference task. We map the stack of tomographic convergence maps, treated as input channels,
to a $d=10$ dimensional summary $\boldsymbol{s} = f_\eta(\boldsymbol{\kappa})$ through a CNN with
weights $\eta$. The network is trained with the VMIM objective, which maximises a variational
lower bound on the mutual information between the summary and the parameters
\citep{jeffrey_dark_2024},
\begin{equation}
\label{eq:vmim}
\eta^\star = \argmax_\eta\, \mathbb{E}_{p(\boldsymbol{\theta},\boldsymbol{\kappa})}
             \big[\log q_\psi\big(\boldsymbol{\theta}\,\big|\,f_\eta(\boldsymbol{\kappa})\big)\big],
\end{equation}
where $q_\psi$ is an auxiliary conditional density (a normalising flow) trained jointly with the
compressor. Maximising Eq.~\eqref{eq:vmim} optimises the summary to retain the parameter-relevant
information in the maps. The auxiliary flow is discarded after training and plays no role in the
inference of Sect.~\ref{sec:inference}. The network mixes all four input channels from its first
convolutional layer, so inter-bin information is available to it by construction. The cross-maps
of Sect.~\ref{sec:crossmaps} are therefore constructed only for the per-channel $\ell_1$-norm,
which is evaluated on each channel individually, and we confirm in Sect.~\ref{sec:results} that they
do not improve the CNN.

The compressor is a residual convolutional network of the ResNet-18 family, mapping the stacked
input channels through global average pooling and a dense head to the $d=10$ summary. We
standardise the summary to zero mean and unit variance before inference. We adopted this
architecture after a search over plain, residual and attention-augmented networks, the summary
dimension and the training budget, and, for the density estimator, the flow family and capacity
(Appendix~\ref{app:exhaustiveness}). We report the best-performing configuration of the search.
The alternatives reach within a few per cent of it. On the many-channel input we avoid standard
batch normalisation, whose statistics, computed over batches that mix cosmologies, remove part of
the cosmological signal and reduce the constraining power. The compressor is trained for
$5\times10^4$ steps, checkpointing on the validation loss.

\subsection{Tomographic cross-maps}
\label{sec:crossmaps}

Most of the convergence signal is in the per-bin auto-maps, but the lensing kernels of the
tomographic bins overlap while their shape noise is independent, so the signals of the bins are
correlated, and their cross-correlations constrain the growth of structure and the
distance--redshift relation. A per-channel statistic such as the $\ell_1$-norm, evaluated on one
map at a time, does not measure this inter-bin information. The standard practice in survey
analyses is to merge the galaxy catalogues of two or more bins and to measure the statistic on the
map built from the combined catalogue \citep{martinet_probing_2021, harnois-deraps_cosmic_2021,
vinciguerra_euclid_2026}. Because the map estimator is linear in the source galaxies, such a
combined map is, pixel by pixel, a count-weighted average of the per-bin maps, noise included. The
practice thus supplies the statistic with \emph{linear} cross-bin channels, which determine the
second-order cross-correlations of each pair but, for any fixed set of combinations, only part of
the higher joint structure (Appendix~\ref{app:bnttheory}). We instead supply \emph{quadratic}
channels, explicit cross-maps that are bilinear in the two maps of each bin pair, alongside the
four auto-maps.

All summaries in the comparison, the CNN included, are computed from the four auto-maps of one
$10\degr$ patch. We therefore build the cross-maps from the two auto-maps of the patch. For
each bin pair we form two complementary cross-maps (Fig.~\ref{fig:data}, middle and bottom rows). The
first is a convolution,
\begin{equation}
\label{eq:crossconv}
\kappa^{ij}_{\rm conv} = \mathcal{F}^{-1}\!\big[\,\mathcal{F}(\kappa_i\,W)\;\mathcal{F}(\kappa_j\,W)\,\big],
\end{equation}
where $\mathcal{F}$ is the fast Fourier transform and $W$ a cosine apodisation window. This is the
flat-sky analogue of the harmonic-space product of spherical-harmonic coefficients used by
\citet{zurcher2022dark, zurcher_towards_2023}. Each pixel of the
convolution map is a sum over all pixel pairs of the patch, and, as we measured on the simulated
maps, this concentrates the map's variance in a few tens of effective large-scale modes. Thus, its
one-point statistics carry mostly two-point cross information. The second is the pointwise product,
\begin{equation}
\label{eq:crossprod}
\kappa^{ij}_{\rm prod} = \kappa_i\,\kappa_j,
\end{equation}
which we introduce as the local counterpart of the convolution. It multiplies the two maps pixel
by pixel, so the structure that the two maps share at the same position enters coherently, and it
preserves the small scales that the convolution suppresses. Its spatial mean is the zero-lag
cross-correlation $\langle\kappa_i\kappa_j\rangle$, and its higher one-point moments are the joint
moments $\langle\kappa_i^n\kappa_j^n\rangle$ of the two maps, that is, non-Gaussian information
between the bins. We consider four input configurations for the $\ell_1$-norm: the auto-maps alone; the
auto-maps with the six convolution cross-maps; with the six product cross-maps; and with both, sixteen
channels in total.

Either cross-map reduces each bin pair to a single field before the statistic is taken, whereas
the joint $\ell_1$-norm of Sect.~\ref{sec:jointl1} is computed from the full two-dimensional
distribution of the pair's wavelet coefficients, scale by scale. Section~\ref{sec:power} compares
the four configurations.
\section{Inference and calibration}
\label{sec:inference}

\begin{figure*}[t]
  \centering
  \includegraphics[width=\textwidth]{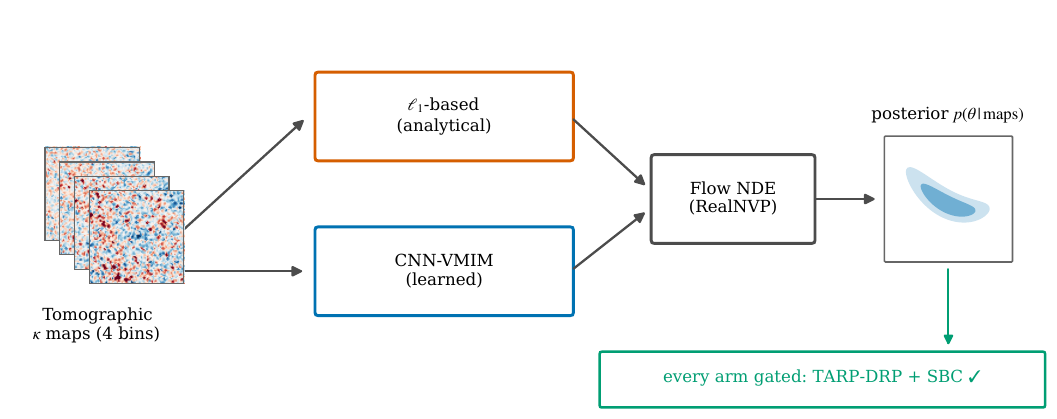}
  \caption{The inference design. Each summary of the tomographic convergence maps, the
  $\ell_1$-norm or the CNN-VMIM compressor, is reduced to a
  ten-dimensional vector and passed through the \emph{same} conditional RealNVP to give the
  parameter posterior, so that any difference between the posteriors comes from the summaries
  alone. Every summary is gated by the calibration tests of Sect.~\ref{sec:calibration}.}
  \label{fig:pipeline}
\end{figure*}

We infer the posterior of the cosmological parameters with neural posterior estimation (NPE), and
we verify the calibration of every reported posterior before interpreting its width. To make the
two summaries comparable, each is compressed to the same low dimension and modelled with the same
normalising flow, and every posterior is validated with the same coverage tests
(Fig.~\ref{fig:pipeline}).

\subsection{Neural posterior estimation}
\label{sec:npe}

Given a summary $\boldsymbol{s}$ of the convergence maps, we estimate the posterior
$p(\boldsymbol{\theta}\,|\,\boldsymbol{s})$ with a conditional normalising flow, trained by
maximising the expected log-probability of the true parameters over the training pairs. Our
density estimator throughout is a coupling-based flow of the RealNVP type
\citep{dinh_density_2017}, with four coupling blocks of width $128$. It is optimised with Adam
under a warmup-then-cosine learning-rate schedule, with early stopping on a held-out validation
split. Each summary's posterior is estimated by pooling three such flows trained from independent
initialisations, which averages the estimator's own training stochasticity. We also tested
masked autoregressive flows \citep[MAF;][]{papamakarios_masked_2017}, which gave slightly weaker
calibrated constraints for both summaries (Appendix~\ref{app:exhaustiveness}).

We optimised the inference setup for each summary independently, over the flow family and
capacity (Appendix~\ref{app:exhaustiveness}), and the same flow configuration proved optimal for
all of them. We chose ten dimensions for the summaries rather than the minimum of six (the number
of parameters we are constraining) since slightly larger summaries gave marginally better
constraints in the search of Appendix~\ref{app:exhaustiveness}. The $\ell_1$ data vector, of
dimension of order $10^3$, is compressed to ten components through a small fully-connected
network of two hidden layers of $256$ units trained with the same variational mutual-information
objective as the CNN, so that both summaries are modelled by the same density estimator with an
identical configuration. We verified that this compression is effectively lossless
(Appendix~\ref{app:exhaustiveness}). The joint $\ell_1$-norm, and every $\ell_1$ summary in the
BNT basis, is inferred through an ensemble of three independently trained compressors whose
posteriors we pool per observation. This averages over the stochasticity of the variational
compression and gives a calibrated posterior (Appendix~\ref{app:calibration}).

\subsection{Calibration tests}
\label{sec:calibration}

Because a learned pipeline can produce uncalibrated posteriors, we subject every posterior to the
same coverage tests before quoting its constraints, and we treat passing them as a precondition
for reporting constraining power. The results are collected in Appendix~\ref{app:calibration}.

The two tests we apply to every summary are the tests of accuracy with random points (TARP), for
the global expected coverage in the full parameter space \citep{lemos_sampling-based_2023}, and
simulation-based calibration (SBC), for the uniformity of the posterior rank statistics over the
validation cosmologies \citep{talts_validating_2020}. The first measures, at each credibility
level, the fraction of true parameters that fall inside the corresponding credible region,
evaluated over parameters drawn from the prior. For a calibrated posterior this fraction lies on
the identity line. The second uses the rank of the true parameter among the posterior samples to
test whether any parameter's posterior is systematically biased or over- or under-dispersed.
\section{Constraining power}
\label{sec:results}
\label{sec:power}

We now compare the summaries of Sect.~\ref{sec:stats}. For each summary we infer a posterior for
$9000$ mock observations and characterise it by the median over observations of the marginal
uncertainties $\sigma(\Omega_{\rm m})$, $\sigma(\sigma_8)$, $\sigma(w_0)$ and of the
three-parameter figure of merit
$\mathrm{FoM}_3 = 1/\sqrt{\det \mathbf{C}_{\Omega_{\rm m},\sigma_8,w_0}}$
(Sect.~\ref{sec:simulations}).

\begin{figure*}[t]
  \centering
  \includegraphics[width=\textwidth]{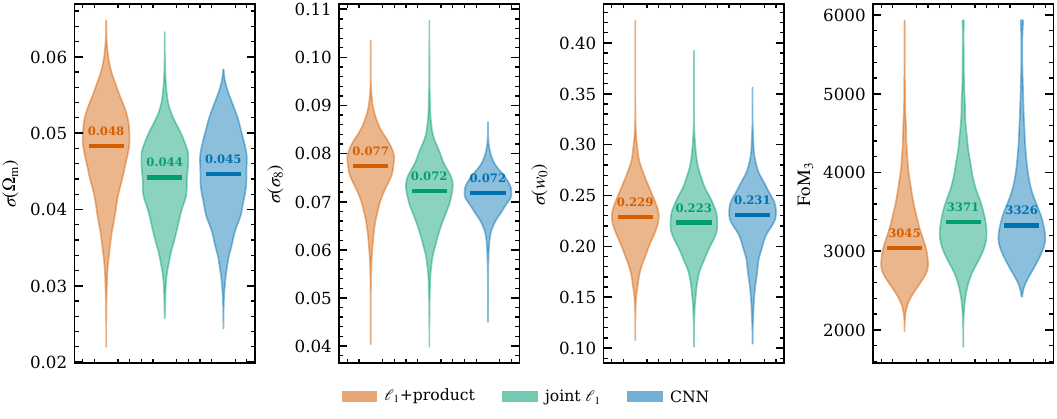}
  \caption{Per-patch distributions of the marginal uncertainties $\sigma(\Omega_{\rm m})$,
  $\sigma(\sigma_8)$, $\sigma(w_0)$ and the three-parameter figure of merit over the $9000$ mock
  observations, for the $\ell_1$+product summary, the joint $\ell_1$-norm,
  and the CNN. The distributions of the joint $\ell_1$-norm and the CNN overlap closely in all four
  panels (medians in Table~\ref{tab:headline}).}
  \label{fig:cp}
\end{figure*}

\begin{figure}[t]
  \centering
  \includegraphics[width=\columnwidth]{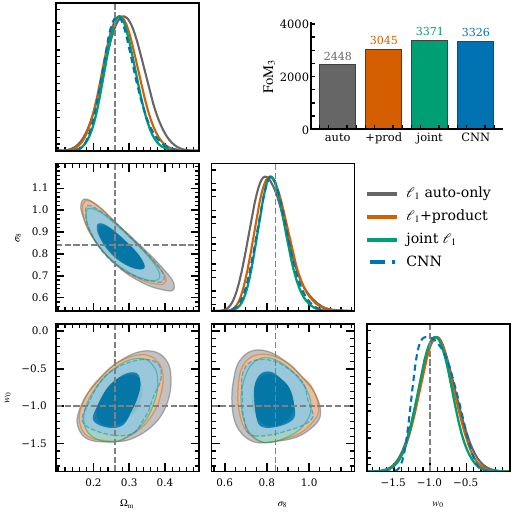}
  \caption{Posterior on $(\Omega_{\rm m},\sigma_8,w_0)$ at the mean observation for the four
  summaries: the per-channel $\ell_1$ auto-only, the $\ell_1$+product, the joint $\ell_1$-norm, and the CNN
  compressor (dashed). The inset bars give the median $\mathrm{FoM}_3$ over the mock observations.
  The auto-only $\ell_1$ contour is the widest, the $\ell_1$+product
  contour approaches the CNN's, and the joint $\ell_1$-norm's nearly coincides with it.}
  \label{fig:corner}
\end{figure}

Table~\ref{tab:headline} compares the summaries, all inferred through the same pipeline
(Sect.~\ref{sec:inference}). On the auto-maps alone the per-channel $\ell_1$-norm reaches a figure of
merit $26\%$ below the CNN's, as expected from Sect.~\ref{sec:stats}. With the product cross-maps
the difference shrinks to $8\%$, with $\sigma(w_0)$ already at the CNN's value ($0.229$ against
$0.231$; Fig.~\ref{fig:cp}), and with the joint $\ell_1$-norm the two figures of merit agree
within the quoted uncertainties. Measured over the $9000$ mock observations, these medians and
uncertainties show that the joint $\ell_1$-norm extracts roughly the same amount of cosmological
information from the maps as the optimised learned compressor. Figure~\ref{fig:corner} shows the
posterior contours at the mean observation, that is, for the summary averaged over the mock
observations, which centres the contours on the true parameters. The joint $\ell_1$ and CNN
contours nearly coincide. The uncertainties quoted on the figure of merit here and in
Sect.~\ref{sec:bnt} are the spread of the median over three independently trained compressors,
which is the dominant source of variability between runs. The training stochasticity of the
density estimator moves the medians by about one per cent or less, and the sampling error of the
median over the finite set of mock observations is below half a per cent.

\begin{table}[t]
\centering
\caption{Constraining power of the summaries. Smaller $\sigma$ and larger $\mathrm{FoM}_3$ mean
stronger constraints. Values are medians over the mock observations of Sect.~\ref{sec:simulations}. The uncertainty on the figure of
merit is the spread over three independently trained compressors (Sect.~\ref{sec:power}).}
\label{tab:headline}
\begin{tabular}{l ccc c}
\hline\hline
Summary & $\sigma(\Omega_{\rm m})$ & $\sigma(\sigma_8)$ & $\sigma(w_0)$ & $\mathrm{FoM}_3$ \\
\hline
$\ell_1$ auto-only     & 0.053 & 0.085 & 0.245 & $2448 \pm 27$  \\
$\ell_1$ +conv         & 0.052 & 0.081 & 0.237 & $2671 \pm 133$ \\
$\ell_1$ +product      & 0.048 & 0.077 & 0.229 & $3045 \pm 183$ \\
$\ell_1$ +conv+product & 0.045 & 0.071 & 0.220 & $3255 \pm 200$ \\
joint $\ell_1$         & 0.044 & 0.072 & 0.223 & $3371 \pm 96$  \\
CNN                    & 0.045 & 0.072 & 0.231 & $3326 \pm 30$  \\
\hline
\end{tabular}
\end{table}

\begin{figure}[t]
  \centering
  \includegraphics[width=0.8\columnwidth]{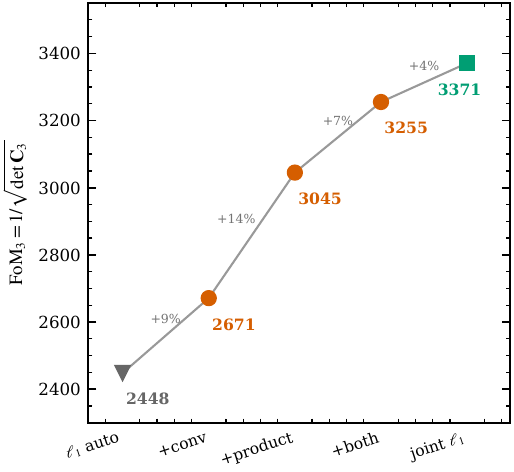}
  \caption{The three-parameter figure of merit of the $\ell_1$-norm as information between bins is
  added: the auto-maps alone, with the convolution cross-maps,
  with the product cross-maps, with both, and the joint $\ell_1$-norm. The labels give the increase
  from one point to the next. The joint $\ell_1$ value is obtained with the three-compressor
  ensemble (Sect.~\ref{sec:npe}). The CNN value is in Table~\ref{tab:headline}.}
  \label{fig:ablation}
\end{figure}

Figure~\ref{fig:ablation} shows how the figure of merit of the $\ell_1$-norm grows as information
between bins is added. Relative to the auto-maps alone, the convolution cross-maps raise it by
$9\%$ and the product cross-maps by $24\%$, so the pixel-wise product is the more informative of
the two cross-map constructions (Sect.~\ref{sec:crossinfo}). With both cross-maps the gain is
$33\%$, close to the sum of the two, so the information they add is largely complementary. The
joint $\ell_1$-norm, computed from the auto-maps alone, gains $38\%$. We verified that explicit
cross-maps do not improve the CNN, as expected since it takes the four maps as joint input
(Sect.~\ref{sec:cnn}).

\section{The BNT transform: contour inflation as an artefact of the basis}
\label{sec:bnt}

The BNT transform \citep{bernardeau_cosmic_2014} is a linear,
invertible recombination of the tomographic bins that localises the lensing kernel in redshift. It
is a change of basis of the tomographic maps, and we call the transformed maps the nulled basis. A
systematic that is localised in redshift then affects only the corresponding nulled maps and scales,
where it can be removed (Sect.~\ref{sec:intro}), which makes the transform attractive as a step
before a non-Gaussian analysis. Applied to higher-order statistics such as the $\ell_1$-norm or
peak counts, however, it has been found to inflate the cosmological contours significantly
\citep{tersenov_baryonic_2026, vinciguerra_euclid_2026}. Using the pipeline of Sect.~\ref{sec:stats}, we show that
this contour inflation is an artefact of the analysis basis, and that summaries which use the
maps jointly are much less affected by it.

\subsection{The contour inflation of each summary}
\label{sec:bnt_result}

\begin{figure*}[t]
  \centering
  \includegraphics[width=0.52\textwidth]{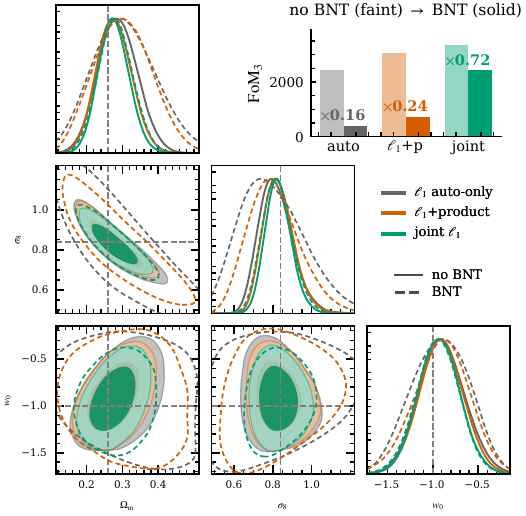}\hfill
  \includegraphics[width=0.45\textwidth]{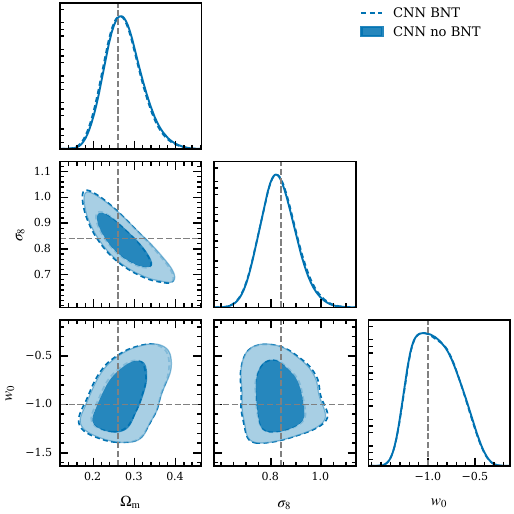}
  \caption{Posteriors on $(\Omega_{\rm m},\sigma_8,w_0)$ without (filled) and with (dashed) the BNT
  transform. \emph{Left:} the analytical summaries ($\ell_1$
  auto-only, $\ell_1$+product, and the joint $\ell_1$-norm). The per-channel statistics widen
  markedly under the transform while the joint $\ell_1$-norm changes little. The inset gives the
  ratio of the figure of merit with and without the transform ($0.16$, $0.24$, $0.72$).
  \emph{Right:} the learned CNN, whose no-BNT and BNT posteriors are near-coincident
  ($\times0.96$).}
  \label{fig:bnt}
\end{figure*}

We apply the BNT transform to the four noisy auto-maps and repeat the analysis with the same
pipeline, with architecture, training and flow identical to the no-BNT case. Only the per-channel
noise normalisation of the $\ell_1$-norm is re-measured on BNT-transformed noise maps, since the
transform changes the noise level of each channel. Table~\ref{tab:bnt} collects the results and
Fig.~\ref{fig:bnt} shows the posteriors. The per-channel $\ell_1$-norm on the auto-maps keeps only $0.16$ of its no-BNT figure of merit. Adding
the cross-maps raises this ratio moderately, to $0.17$ with the convolution and $0.24$ with the
product. The joint $\ell_1$-norm keeps $0.72$ of its no-BNT figure of merit, and the CNN, which
takes the four maps as joint input and can therefore use the correlations between them by
construction, is near-lossless under BNT, with marginal uncertainties that change by at most a few
per cent. Section~\ref{sec:bnt_whiten} explains why the summaries that use the maps jointly are
less affected by BNT, even though the nulling is designed to de-correlate the signal between
tomographic bins.

\begin{table}[t]
\centering
\caption{BNT contour inflation: the three-parameter figure of merit without and with the BNT
transform, and their ratio (BNT\,/\,no-BNT). All values are calibrated through the same coverage
tests (Appendix~\ref{app:calibration}). The $\ell_1$ summaries in the BNT basis and the joint
$\ell_1$-norm in both bases are obtained with the three-compressor ensemble of
Sect.~\ref{sec:npe}, the others with a single compressor. The uncertainty on each figure of merit
is the spread over three independently trained compressors (Sect.~\ref{sec:power}).}
\label{tab:bnt}
\setlength{\tabcolsep}{4pt}
\begin{tabular}{l ccc}
\hline\hline
Summary & no BNT & BNT & BNT/no-BNT \\
\hline
$\ell_1$ auto-only     & $2448 \pm 27$  & $388 \pm 43$   & 0.16 \\
$\ell_1$ +conv         & $2671 \pm 133$ & $458 \pm 15$   & 0.17 \\
$\ell_1$ +product      & $3045 \pm 183$ & $718 \pm 29$   & 0.24 \\
$\ell_1$ +conv+product & $3255 \pm 200$ & $704 \pm 76$   & 0.22 \\
joint $\ell_1$         & $3371 \pm 96$  & $2424 \pm 208$ & 0.72 \\
CNN                    & $3326 \pm 30$  & $3186 \pm 52$  & 0.96 \\
\hline
\end{tabular}
\end{table}

\subsection{The contour inflation is an artefact of the basis}
\label{sec:bnt_whiten}

The SNR of the individual maps is much lower in the BNT basis compared to the original one.
In the original basis each map has a broad lensing kernel, which integrates the signal over a long stretch of the line of sight, while in the BNT basis the maps have thin kernels (except for the first tomographic map, which remains unchanged), so they individually contain much less signal. 
Their noise level, at the same time, is higher: each nulled map is a linear combination of several original maps, whose shape noises are independent, so the noise variances add up. 
This is why a statistic evaluated on each map individually (such as the per-channel $\ell_1$-norm) loses constraining power \citep{tersenov_baryonic_2026}.
We verified that with noiseless maps the BNT and no-BNT contours essentially coincide for every summary considered here. 

The transform, however, is linear and invertible, so the four maps together must contain the same
information in both bases. Since each map on its own now carries much less, the difference lies in
the combination of the maps, and a statistic has to capture the cross-map information to recover
it. The auto- and cross-spectra, for example, transform among themselves invertibly
(Appendix~\ref{app:bnttheory}), and two-point analyses that use them are lossless under the
nulling \citep{vinciguerra_euclid_2026, tersenov_baryonic_2026}. Table~\ref{tab:bnt} shows the
same dependence for the higher-order summaries. The ratio of the figure of merit with and without the
transform increases in the order in which these summaries capture more of the cross-map
information (Sect.~\ref{sec:stats}). The joint $\ell_1$-norm keeps most of its constraining power,
whereas the compressor is near-lossless. The reason for that is that the joint
$\ell_1$-norm captures the distribution of the coefficients of only two maps at a time, so it misses
the structure that involves three or four maps at once. This structure contributes little
constraining power in the original basis, where the pairwise statistic is already as constraining as the compressor
(Sect.~\ref{sec:results}), but
in the BNT basis it carries a much larger share of the constraining power. A joint $\ell_1$-norm
over all four bins would in principle be as insensitive to the transform as the compressor
(Appendix~\ref{app:bnttheory}), but its four-dimensional histogram is too sparse to populate on a
patch (Sect.~\ref{sec:jointl1}).
\section{Discussion}
\label{sec:discussion}

\subsection{What the cross-maps contribute}
\label{sec:crossinfo}

Adding the cross-maps to the auto-maps raises the figure of merit of the $\ell_1$-norm by up to
$33\%$, most of it through the product cross-maps (Sect.~\ref{sec:power}). We attribute the
difference to the locality of the two constructions. The convolution map is a sum over the whole
patch and retains only a few tens of effective modes, whereas the product map is formed pixel by
pixel and is sensitive to the small-scale structure that the two maps share
(Sect.~\ref{sec:crossmaps}). For comparison, the larger cross-bin improvements reported for
contiguous survey footprints come largely from the self-calibration of intrinsic alignments
\citep{zurcher_towards_2023}, which our forward model does not include.

\subsection{Why the starlet $\ell_1$-norm can be as constraining as a learned compressor}
\label{sec:whytie}

The fact that our fixed, analytical statistic is as constraining as the optimised neural
compressor reflects the simplicity of the convergence maps at the scales and
depths of this analysis. To a good approximation, a noisy convergence map is a Gaussian random
field plus quasi-circular peaks from collapsed structures, and much of the complex morphology
beyond this lies below the shape noise. Additional evidence for this comes from the
map-reconstruction problem, where a mass-mapping method built on exactly this two-component
model, a Gaussian field plus a sparse wavelet component \citep[MCALens;][]{starck_weak-lensing_2021},
was shown to perform comparably to deep generative reconstructions \citep{remy_probabilistic_2023}.
There is therefore little structure in the maps for a learned compressor to exploit beyond this.

The starlet $\ell_1$-norm is well suited to this structure of the convergence maps. Its isotropic multiscale atoms
resemble the peaks in the field, and the statistic is computed from the full amplitude
distribution of the wavelet coefficients scale by scale, Gaussian core and non-Gaussian tails
alike \citep{ajani_starlet_2021}. Accordingly, in the Euclid comparisons of higher-order
statistics the multiscale $\ell_1$-norm is among the best-performing statistics
\citep{ajani_forecasts_2023, vinciguerra_euclid_2026}. Where the data are morphologically richer
than the weak-lensing maps studied here, a learned compressor has structure to exploit that no
simple fixed statistic encodes.

\subsection{Analytical versus learned summaries}
\label{sec:analyticalvslearned}

The two summaries are equally constraining within the uncertainties, so the choice between them
depends on their other properties. The $\ell_1$-norm is a simple, fixed function of the maps, so it requires no
training, is inexpensive to compute, and is identical in every reanalysis. Its data vector can be
inspected directly as a function of scale and SNR, so an anomaly in a measured data
vector can be traced to the scales and thresholds that produce it. On the larger scales and higher
source redshifts where the field is only weakly non-linear, the per-channel statistic can moreover
be predicted analytically \citep{sreekanth_theoretical_2024}. No such prediction exists yet for
the joint $\ell_1$-norm.

The learned compressor is optimised for the parameters and simulations at hand. It combines the
tomographic bins by construction, which makes it near-lossless under the BNT transform, where
every fixed statistic considered here loses constraining power (Sect.~\ref{sec:bnt}). It must be retrained, its convergence
verified and its posterior recalibrated for every change of survey configuration or forward model.
And since it is optimised for information on the simulations, it cannot distinguish physical
features from features of the simulator, such as its resolution, noise model or missing physics,
and is therefore more exposed than a fixed statistic to features that are absent from real data. A fixed statistic is also
affected by forward-model error, but it measures only the features it was designed for, and a
mismatch can sometimes be recognised in the data vector itself.

\subsection{Using the nulled basis in practice}
\label{sec:bntpractice}

The practical message of Sect.~\ref{sec:bnt} is that the nulling need not cost a higher-order
analysis its constraining power, provided some stage of the pipeline uses the tomographic maps
jointly. Combining the galaxy catalogues of several bins before map-making, the standard
construction of cross-statistics (Sect.~\ref{sec:crossmaps}), is not sufficient. The combined map
is a linear combination of the per-bin maps, so a statistic measured on it is still evaluated on
a few fixed linear combinations of the maps, and the contour inflation persists, as observed for
the combined-bin higher-order statistics of \citet{vinciguerra_euclid_2026} under the nulling. An
alternative is to keep the statistics out of the nulled basis altogether: apply the transform,
impose the scale cuts in the nulled basis, where a systematic is confined to known channels and
scales, and invert the transform, so that the statistics are measured on cleaned maps with broad
kernels \citep{gu_mitigating_2025, vinciguerra_euclid_2026}. Since the cut differs between the
nulled channels, however, the cut and the inversion do not commute exactly. The maps returned to
the original basis are unchanged on the scales that pass every cut and empty on the scales that
pass none, but on the scales in between each returned map is a fixed linear mixture of the
original maps, with shape noise correlated between them. On these scales a per-channel statistic
loses constraining power for the same reason as in the nulled basis, in milder form. How much
this costs for realistic cuts remains to be quantified.

The problem does not arise for the learned compressor, since a joint summary can absorb any fixed invertible
recombination of its input channels in its first layer (Appendix~\ref{app:bnttheory}). Another
possibility, which we have not explored, is to combine the maps at the map-making stage. Since
the per-channel loss is an SNR effect (Sect.~\ref{sec:bnt_whiten}), a mass-mapping
method that reconstructs and denoises the tomographic maps jointly, such as a tomographic
extension of the plug-and-play reconstruction of \citet{leterme_plug-and-play_2026}, could act
directly in the nulled basis and deliver nulled maps of low enough noise for a per-channel
statistic to keep its constraining power. Both routes keep the analysis in the basis where a
residual systematic remains localised.

\subsection{Limitations and outlook}
\label{sec:limitations}

One objection that could be raised against the comparison of Sect.~\ref{sec:power} is that the
learned side might not be
fully optimised, so that the comparison understates the potential performance of a learned
compressor. To address this, we varied the architecture, the summary dimension, the training
budget and the inference flow (Appendix~\ref{app:exhaustiveness}) and retained the best
configuration of each. The alternatives change the constraining power by no more than the
seed-to-seed scatter. A second possible objection is data scarcity. The simulation set is large
(Sect.~\ref{sec:simulations}) and both summaries are built from the same simulations, but we
cannot exclude that a denser sampling of parameter space would benefit the learned compressor
more than the analytical summary. The compressor keeps a measurable advantage only under the BNT
transform, where the figure of merit of the joint $\ell_1$-norm drops to $0.72$ of its no-BNT
value while the compressor is near-lossless (Sect.~\ref{sec:bnt_whiten}). A summary that captures the structure of three or
four maps at once, such as a compact set of higher joint cumulants, would be the natural way to
close this difference.

A caveat on the metric is that the three-parameter figure of merit depends on the correlation of
the posterior, which is strong in the $(\Omega_{\rm m},\sigma_8)$ plane, so we report it
alongside the marginal uncertainties throughout (Appendix~\ref{app:pitfalls}). Finally, the
analysis is restricted to $10\degr$ patches, on which the flat-sky projection residuals are small
(Sect.~\ref{sec:simulations}) and shared by both summaries.

\section{Conclusions}
\label{sec:conclusions}

We have compared the starlet $\ell_1$-norm with a learned CNN-VMIM compressor for tomographic
weak-lensing inference on $10\degr$ patches, and introduced the joint $\ell_1$-norm, an extension of
the statistic to pairs of redshift bins. Each summary was compressed to a common dimension,
modelled with the same normalising flow, and every posterior verified to be calibrated.
Our main findings are the following.

\begin{enumerate}
  \item \textbf{A fixed analytical statistic is as constraining as the optimised learned
        compressor.} With the product cross-maps as extra channels, the per-channel $\ell_1$-norm
        comes within $8\%$ of the compressor's figure of merit and reaches its $\sigma(w_0)$. The
        joint $\ell_1$-norm reaches the same figure of merit as the compressor within the uncertainties
        ($\mathrm{FoM}_3 = 3371 \pm 96$ against $3326 \pm 30$). We interpret this as both
        summaries extracting roughly all the cosmological information accessible in the maps
        (Sect.~\ref{sec:whytie}).
  \item \textbf{The information between bins accounts for the difference between the per-channel
        $\ell_1$-norm and the compressor.} On the
        auto-maps alone the per-channel $\ell_1$-norm reaches a figure of merit $26\%$ below the
        compressor's, which takes the four bins as joint input. Two constructions supply the missing information: the product
        cross-maps recover most of the difference, and the joint $\ell_1$-norm the rest. The
        convolution cross-maps raise the figure of merit by $9\%$, against $24\%$ for the product
        cross-maps, and explicit cross-maps do not improve the compressor.
  \item \textbf{The BNT contour inflation is an artefact of the basis.} Under the BNT transform the
        figure of merit of the per-channel $\ell_1$-norm drops to $0.16$ of its no-BNT value ($0.24$
        with the product cross-maps), that of the joint $\ell_1$-norm to $0.72$, and the CNN is
        near-lossless. The
        transform is linear and invertible, so no information is lost. But each nulled map on its
        own carries little signal, and the transform correlates the shape noise across the maps, so
        the constraint can be recovered only from the maps jointly. A statistic evaluated on each map
        individually cannot do this. This explains the inflation reported by
        \citet{tersenov_baryonic_2026} and gives a criterion for which summaries lose constraining
        power under the nulling (Sect.~\ref{sec:bnt} and Appendix~\ref{app:bnttheory}).
\end{enumerate}

For constraining power, a learned compressor of the weak-lensing maps is thus not needed. A fixed analytical statistic, once
given the information between bins, extracts as much from the tomographic maps as the optimised
compressor, without training and with a data vector that can be inspected directly
(Sect.~\ref{sec:analyticalvslearned}). The remaining advantage of the neural compressor is that it
combines the bins by construction, so it needs no cross-maps and keeps its constraining power in
the nulled basis. A practical consequence is that the BNT transform can be kept as a mitigation
step for redshift-localised systematics such as baryonic feedback at little cost in constraining
power if the summary uses the maps in pairs, as the joint $\ell_1$-norm does, and at no cost if it
uses all of them at once, as the CNN does.
\section*{Data availability}

The \texttt{cosmoGRID\,V1} simulations are publicly available \citep{kacprzak_cosmogridv1_2023}.
The analysis code and the summary data vectors will be made available on GitHub upon
publication.

\begin{acknowledgements}
This work was funded by the TITAN ERA Chair project (contract no. 101086741) within the Horizon
Europe Framework Program of the European Commission, and the French Agence Nationale de la
Recherche (ANR-22-CE31-0014-01 TOSCA).
We thank François Lanusse for helpful discussions. We also acknowledge the developers of the \texttt{cosmoGRID} simulation suite for making their data publicly available.
\end{acknowledgements}

\bibliographystyle{aa}     %
\bibliography{refs,new_astro_bib,pending_bib_additions}   %

\appendix
\section{Calibration}
\label{app:calibration}

This appendix collects the calibration results behind every constraint in the paper. We gate every
summary, in both the standard and the BNT basis, with the two coverage tests of
Sect.~\ref{sec:calibration}, TARP for the global expected coverage and SBC for the uniformity of
the posterior rank statistics, both evaluated over parameters drawn from the prior. Both tests use $N=600$ validation points, which resolve coverage
deviations down to a binomial band of ${\sim}0.02$. The calibration statements in this paper hold
at that resolution. The figures below show the $\ell_1$ auto-only, the $\ell_1$+product, the joint
$\ell_1$-norm and the CNN. The $\ell_1$+conv and $\ell_1$+conv+product summaries pass the same tests
(not shown).

\begin{figure}[t]
  \centering
  \includegraphics[width=0.8\columnwidth]{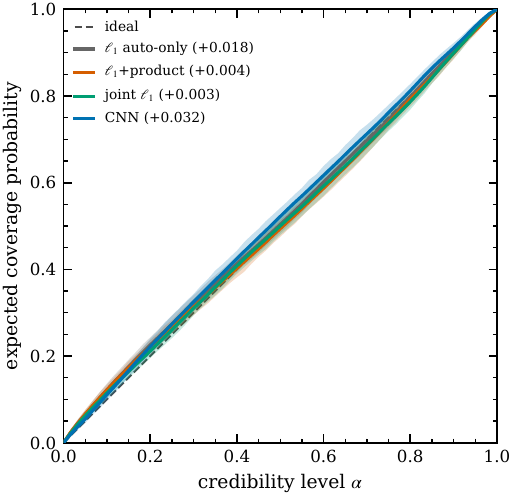}
  \caption{Expected-coverage (TARP) test in the standard basis for the four summaries. The grey band is the $1\sigma$ uncertainty from a bootstrap over the
  $N=600$ validation sightlines, and the legend gives the net coverage (positive = conservative).
  All summaries lie near the diagonal.}
  \label{fig:tarp}
\end{figure}

\begin{figure}[t]
  \centering
  \includegraphics[width=0.8\columnwidth]{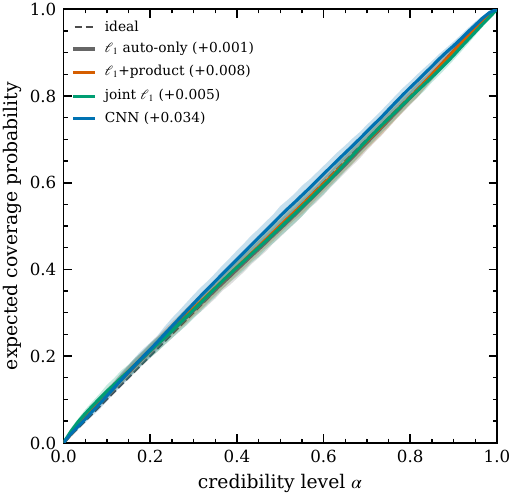}
  \caption{As Fig.~\ref{fig:tarp}, but in the BNT basis, with the analytical summaries as the
  three-compressor ensembles. All summaries lie near the diagonal.}
  \label{fig:tarp_bnt}
\end{figure}

\paragraph{Expected coverage.}
Figures~\ref{fig:tarp} and~\ref{fig:tarp_bnt} show the TARP expected-coverage curves in the
standard and the BNT basis. The grey band is the $1\sigma$ uncertainty from a bootstrap over the
$N=600$ validation sightlines (the binomial standard error $\sqrt{0.25/N}\approx0.02$ at the
median). The integrated net coverage (positive means conservative, negative over-confident) is, in
the standard basis, $+0.018$, $+0.004$, $+0.003$ and $+0.032$ for the $\ell_1$ auto-only, the
$\ell_1$+product, the joint $\ell_1$-norm and the CNN. In the BNT basis it is $+0.001$, $+0.008$,
$+0.005$ and $+0.034$. The analytical summaries are consistent with ideal coverage to better than
$0.5\sigma$ in both bases. The CNN is mildly conservative ($\sim\!1.4\sigma$, not significant).
Under BNT the analytical summaries, gated as three-compressor ensembles, remain calibrated despite
their inflated contours. Their single-compressor BNT posteriors were mildly over-confident, and
the three-compressor ensemble, which restores calibration, is the one used throughout.

\begin{figure*}[t]
  \centering
  \includegraphics[width=0.9\textwidth]{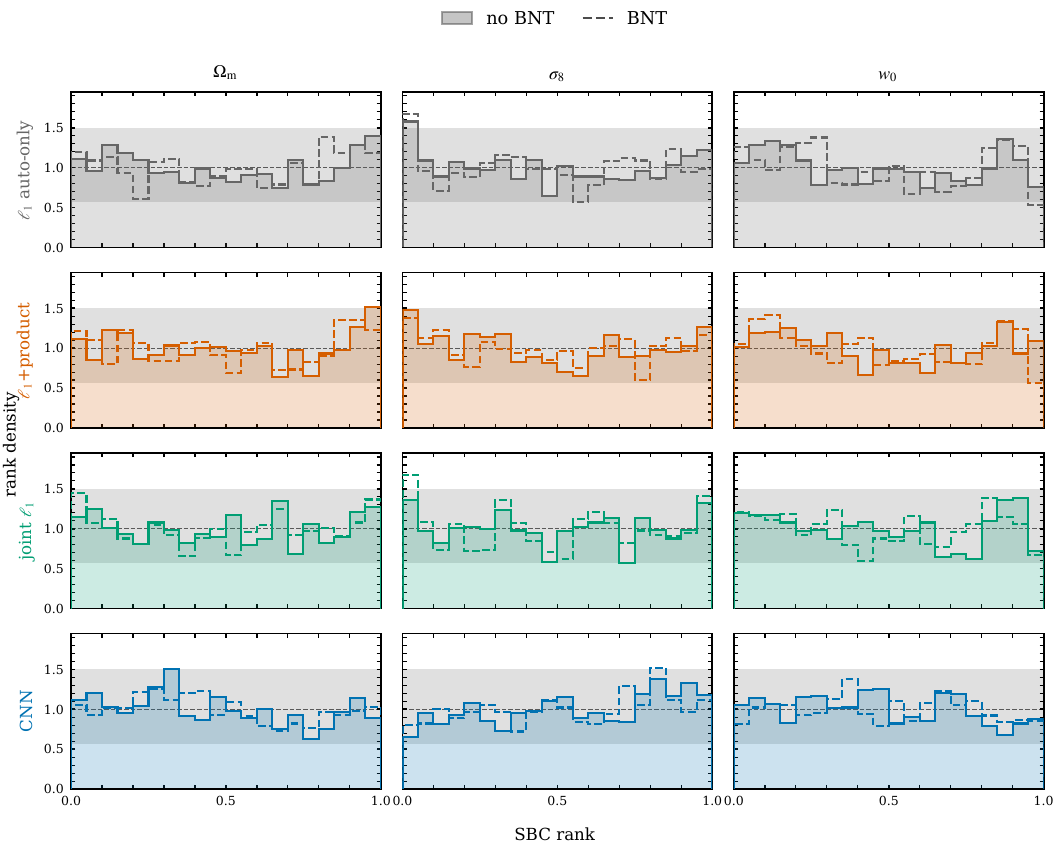}
  \caption{SBC rank histograms for the four summaries (rows) and the three science
  parameters (columns), with the standard basis (filled) and the BNT basis (dashed outline) overlaid in each
  panel. Bars are the mean over three independently trained flows. The grey band is the central $99\%$ of
  the binomial scatter expected under uniformity for the $N=600$ independent test points. A $\cup$ shape
  indicates over-confidence, an $\cap$ shape under-confidence, and a slope a biased mean. All summaries lie
  within the band in all three parameters and both bases.}
  \label{fig:sbc}
\end{figure*}

\paragraph{Rank statistics.}
Figure~\ref{fig:sbc} gives the SBC rank histograms for the four summaries (rows) and the three
science parameters (columns), with the standard basis (filled) and the BNT basis (dashed) overlaid
in each panel. The rank is the fraction of posterior samples below the truth at each prior-drawn
test point and is uniform under calibration. A $\cup$ shape signals over-confidence, an $\cap$
shape under-confidence, and a slope a biased mean. The grey band is the central $99\%$ of the
binomial scatter expected under uniformity for the $N=600$ independent test points. The three
retraining seeds share identical test parameters, so the number of independent test points is
$600$ per seed, and the bars are the mean over the three seeds. The rank standard deviations in
the standard basis are $0.305$, $0.302$, $0.298$ and $0.290$ against the uniform value $0.289$, a
mild over-confidence at the few-per-cent level. All summaries lie within the band in all three
parameters, in both bases.

\section{Validation and robustness}
\label{app:pitfalls}

The comparison in the main text rests on several preprocessing and validation choices, and we
record here the checks that justify them.

\paragraph{Mean subtraction.}
A convergence map carries an additive offset that a real survey cannot measure, owing to the
mass-sheet degeneracy. We subtract the mean of each channel before compression. Without this step
a learned compressor treats the offset as an informative feature and reports constraints far
tighter than the data can support. The $\ell_1$-norm is insensitive to the offset, since the
wavelet filters are compensated, but we apply the subtraction to all summaries for uniformity
(Sect.~\ref{sec:simulations}).

\paragraph{Per-channel noise model.}
The cross-maps are smaller in amplitude than the auto-maps by several orders of magnitude. With a
single noise level shared by all channels, the wavelet coefficients of the cross-maps would have
SNR close to zero, so that their histograms would be confined to the central SNR bins and carry
no inter-bin information. We therefore set the noise level per channel from a fixed estimate, and
we verified that the cross-channel histograms then span the SNR range (Sect.~\ref{sec:l1}).

\paragraph{The figure of merit.}
The three-parameter figure of merit is sensitive to the correlation of the posterior. With
$\rho(\Omega_{\rm m},\sigma_8)\approx-0.93$, a change of a few per cent in the marginal
uncertainties moves $\mathrm{FoM}_3$ by tens of per cent. We therefore quote it alongside the
marginal uncertainties throughout.
\section{Optimisation and robustness of the comparison}
\label{app:exhaustiveness}

This appendix summarises the optimisation of the learned pipeline (the compressor architecture,
its training, and the inference flow) and the robustness of the comparison to these choices.

\subsection{Training set and architecture}
\label{app:ceiling}

The compressor is trained on the $3.2\times10^5$ patch examples from $899$ cosmologies of
Sect.~\ref{sec:simulations}, the same simulations on which the density estimators of both summaries
and the compression of the $\ell_1$ data vector (Sect.~\ref{sec:npe}) are trained. This is sufficient for feature learning, and the limiting
resource is the number of distinct cosmologies, which is the same for both summaries. We searched
over architecture, from plain convolutional networks through residual networks up to a ResNet-50
with group normalisation and a cross-channel attention block, over the summary dimension, and over
the optimiser schedule and training budget, the latter through a dedicated \texttt{optuna}
hyperparameter search.
The constraining power is stable across these choices to within the seed-to-seed scatter. A deeper
ResNet-50 trained longer reduces the figure of merit and overfits the $899$ cosmologies, its
validation loss reaching its minimum at about a third of training and drifting upward thereafter,
which indicates that the constraint is limited by the sampling of cosmologies. We adopt the
18-layer residual network (ResNet-18), the best-performing architecture. The alternatives, through
the same inference flow, reach within a few per cent of it.

\subsection{The density estimator}
\label{app:ndeswap}

We optimised the inference flow for each summary independently. The search covered the flow
family (the RealNVP against MAFs of several depths and widths, in our own
JAX implementation) and its capacity, including deeper and wider RealNVP variants and a neural
spline flow. The RealNVP of Sect.~\ref{sec:npe} gave the best
calibrated constraints for every summary, and added capacity or flexibility widened the
posteriors. We also verified that modelling the raw $\ell_1$ data vector directly, with a flow
adapted to its dimension, gives the same constraints as the compressed pipeline
(Sect.~\ref{sec:npe}).

\subsection{Robustness of the result}
\label{app:robustness}

The agreement between the joint $\ell_1$-norm and the CNN is stable across the architecture and
estimator choices above, and every reported posterior passes the same calibration tests. Across
three independently trained compressors the two agree within the quoted spread.

\section{Why the BNT contour inflation is an artefact of the basis}
\label{app:bnttheory}

The BNT transform acts pixel by pixel as a fixed, invertible linear map $\boldsymbol{x}' = B\boldsymbol{x}$
on the vector of tomographic values. The matrix is built once, at a fiducial cosmology, and applied to
data and simulations alike. Its weights depend on cosmology only weakly, through ratios of comoving
distances, and away from the construction cosmology the nulled kernels are only approximately compact.
We collect here the formal statements behind Sect.~\ref{sec:bnt}. They make precise why a
per-channel statistic loses constraining power under the transform while a summary that uses the
channels jointly does not.

\paragraph{The transform removes no information.}
For data transformed by a fixed invertible $B$, the change of variables gives a likelihood
$p(\boldsymbol{x}'\,|\,\boldsymbol{\theta}) = p(\boldsymbol{x}\,|\,\boldsymbol{\theta})\,/\,|\det B|^{N}$,
with a Jacobian independent of $\boldsymbol{\theta}$ that cancels in Bayes' theorem, and the Fisher
information is unchanged as well. The posterior is therefore identical in the two bases, and any
contour inflation under BNT is a property of the summary statistic.

\paragraph{A criterion for summaries unaffected by the transform.}
A fixed summary $T$ is unaffected by the transform exactly when its value on the transformed maps
determines its value on the original maps and vice versa, that is, when the map
$T(\boldsymbol{x}) \mapsto T(B\boldsymbol{x})$ is invertible. The full set of auto- and cross-spectra
satisfies this. Writing $C(\ell)$ for the matrix of auto- and cross-spectra at multipole $\ell$,
the transform acts on it as $C(\ell) \mapsto B\,C(\ell)\,B^{\rm T}$, which is inverted by applying
$B^{-1}$ and $B^{-\rm T}$. The posteriors of an auto-plus-cross two-point analysis are therefore
identical in the two bases for any field. The auto-spectra alone do not satisfy the criterion.
They keep only the diagonal of $B\,C(\ell)\,B^{\rm T}$, four numbers that depend on all ten
originals and cannot be inverted without the off-diagonal terms, which the cross-spectra supply.
The per-channel $\ell_1$-norm does not satisfy it either. In the nulled basis it is a histogram of
each channel taken after the channels have been mixed, and marginalising after mixing is
irreversible. The between-channel structure, which is where the information lies in the nulled
basis (Sect.~\ref{sec:bnt_whiten}), is discarded, and no processing of the per-channel histograms
can recover it.

\paragraph{Why the learned compressor is unaffected.}
A network that mixes its input channels forms, in its first layer, linear combinations
$\sum_i K_{oi}\,x_i$ of its inputs. Fed with $B\boldsymbol{x}$, the same layer computes
$\sum_j\big(\sum_i K_{oi}B_{ij}\big)x_j$, that is, the same combinations with kernels $KB$. The
map $f \mapsto f\circ B$ is thus a bijection of the network class onto itself, and the best
achievable information is identical in the two bases. Undoing the nulling is one configuration of
the first layer, available at no capacity cost and before any non-linearity acts. The small
measured departure from a ratio of one (Table~\ref{tab:bnt}) is then an optimisation residual.
Every per-channel statistic is a function of the joint one-point distribution of the four
coefficients at a pixel, and in the nulled basis this joint distribution is the image of the
original under $B$. How much of a summary's constraining power the transform preserves is
therefore set by how much of this joint distribution the summary captures. The one-dimensional
marginals of a per-channel statistic preserve least, the two-dimensional pairwise distributions of
the joint $\ell_1$-norm more, and the full four-channel distribution that the compressor can learn
is exactly invariant. This is the ordering of the ratios in Table~\ref{tab:bnt}
(Sect.~\ref{sec:bnt_whiten}).

\paragraph{Survey practice as linear sampling.}
The same framework covers the practice of measuring per-channel statistics on combined-catalogue
maps (Sect.~\ref{sec:crossmaps}). A combined map is a linear combination
$\boldsymbol{w}\cdot\boldsymbol{\kappa}$ of the bin maps, and its cumulants are polynomials in the
weights. At order $k$, the $k-1$ mixed cumulants of a pair enter through such a polynomial whose
pure terms are fixed by the auto-maps, so combinations at $k-1$ distinct weight ratios determine
them, and mixed cumulants of three bins require three-bin combinations. By the Cram\'er--Wold
theorem, the one-point distributions of \emph{all} linear combinations determine the full joint
distribution. The same combinations can be formed in either basis, since
$\boldsymbol{w}\cdot B\boldsymbol{\kappa} = (B^{\rm T}\boldsymbol{w})\cdot\boldsymbol{\kappa}$.
Pairwise equal-weight combinations, the common practice, are thus complete at second order and
incomplete at third order and beyond. They are a finite linear sampling of the same joint
distribution whose pairwise part the joint $\ell_1$-norm captures in full.

\end{document}